\documentstyle[12pt,epsf,psfig]{article}

\def\beq{\begin{equation}}
\def\eeq{\end{equation}}
\def\bea{\begin{eqnarray}}
\def\eea{\end{eqnarray}}

\begin{document}

\thispagestyle{empty}

\font\fortssbx=cmssbx10 scaled \magstep2
\hbox to \hsize{
      \hfill$\vtop{
\hbox{MADPH-02-1315}
}
$}

\vspace{.5in}

\begin{center}
{\large\bf KamLAND and solar neutrino data eliminate the LOW solution}\\
\vskip 0.4cm
{V. Barger$^1$ and D. Marfatia$^2$}
\\[.1cm]
$^1${\it Department of Physics, University of Wisconsin, Madison, WI
53706, USA}\\
$^2${\it Department of Physics, Boston University, Boston, MA 02215, USA}\\
\end{center}

\vspace{.5in}

\begin{abstract}                                    
The KamLAND reactor antineutrino experiment has detected a $3.4\sigma$ 
flux suppression relative to the expectation if no neutrino 
oscillations occur. We combine KamLAND data with solar neutrino data 
and show that the LMA solution is the only
viable oscillation 
solution to the solar neutrino problem at the $4.4\sigma$ C.~L.    

\end{abstract}

\thispagestyle{empty}
\newpage

The neutral-current measurement at SNO convincingly demonstrated that 
electron neutrinos from the sun undergo a flavor transformation. Yet, the
cause of this conversion was debatable. With the results from the KamLAND
experiment~\cite{kamland}, one
can confidently state that {\it the solar neutrino problem is solved}. All
explanations of the solar anomaly other than that neutrinos oscillate because
they are massive are now either discarded or are sub-leading effects. 
From solar neutrino data alone, it has been deduced that the
Large Mixing Angle (LMA) and LOW solutions are the most likely oscillation
solutions~\cite{Barger:2002iv}. 
Reactor antineutrino data from KamLAND prove that neutrinos oscillate 
with parameters confined to the Large Mixing Angle (LMA) region at the 
$3.4\sigma$ C.~L. We assess how much more stronger this evidence becomes 
when KamLAND's data is combined with solar neutrino data.

Since solar neutrino experiments and the KamLAND experiment have  
different neutrino sources, their systematics are uncorrelated and their 
results independent. A statistical analysis involving a combination of these
two types of experiments entails two distinct analyses, one of the solar 
data and one of KamLAND data. Subsequently the $\chi^2$ contributions of the
two are simply summed. For details and results of the solar analysis used 
in this work, we refer the reader to Ref.~\cite{Barger:2002iv}.
 Here, we briefly describe our 
analysis of the KamLAND data only.

Electron antineutrinos from 20 nuclear reactors in Japan and S.~Korea are 
incident at the KamLAND detector. About 95\% of the unoscillated flux 
originates with baselines between $80-344$ km. We therefore evaluate
the survival probability of the neutrinos in the vacuum limit of two-flavor
oscillations; the transition probability of muon to electron 
neutrinos 
is known to be small at the atmospheric neutrino oscillation 
scale~\cite{CHOOZ}. We use
the spectra from the fission products of $^{235}\rm{U}$, $^{239}\rm{Pu}$, 
$^{238}\rm{U}$ and  $^{241}\rm{Pu}$ provided in Ref.~\cite{nuc}.
We adopt the time-averaged relative fission yields from the fuel components
as provided by the KamLAND collaboration~\cite{kamland}.  
This serves as a good representation of the averaging of time-evolution
effects of the isotope evolution since all the reactors will not start 
and end their cycles at the same times. 
We assume that the fluctuations in 
the power output of each reactor arising from dead time for maintenance 
and seasonal variations of power requirements
 average so that the live times and efficiencies 
of all the reactors are the same. For the inverse neutron $\beta$-decay
process via which antineutrinos are detected, we adopt the cross-section
with nucleon recoil corrections. To determine the expected signal at KamLAND
from each reactor,
the fluxes are convoluted with the survival probability corresponding to the
baseline of the reactor, the 
antineutrino cross-section and the detector response function (with energy
resolution, 7.5\%$/\sqrt{E({\rm{MeV}})}$, and prompt energy threshold at
2.6 MeV~\cite{kamland}). Finally, 
the cumulative expected signal is obtained by summing over all the reactors. 

To evaluate the statistical significance of an oscillation solution, we define
$\chi^2=\chi^2_{\odot}+\chi^2_{\rm {KamLAND}}$, where $\chi^2_{\odot}$ is
defined by Eq.~(9) of Ref.~\cite{Barger:2002iv}, and~\cite{pdg} 
\begin{equation}
\chi^2_{\rm {KamLAND}}=\sum_{i=1}^{8}
2\,(\alpha N_i^{th}-N_i^{exp}+N_i^{exp} \ln {N_i^{exp} \over \alpha N_i^{th}})
+\sum_{i=9}^{13}
2\,\alpha N_i^{th}
+\big({1-\alpha \over \sigma}\big)^2 \,.
\end{equation}
Here, $N_i^{th}$ and $N_i^{exp}$ are the
theoretical and experimental numbers of events in the $i^{\rm {th}}$ bin
(each of width 0.425 MeV) and
$\sigma=6.42$\% is the uncertainty in the event rate 
calculation~\cite{kamland}. The
normalization factor $\alpha$ is allowed to float so as to yield the smallest
$\chi^2_{\rm {KamLAND}}$ for a given set of oscillation parameters. 

\begin{figure}[ht]
\centering\leavevmode
\mbox{\psfig{file=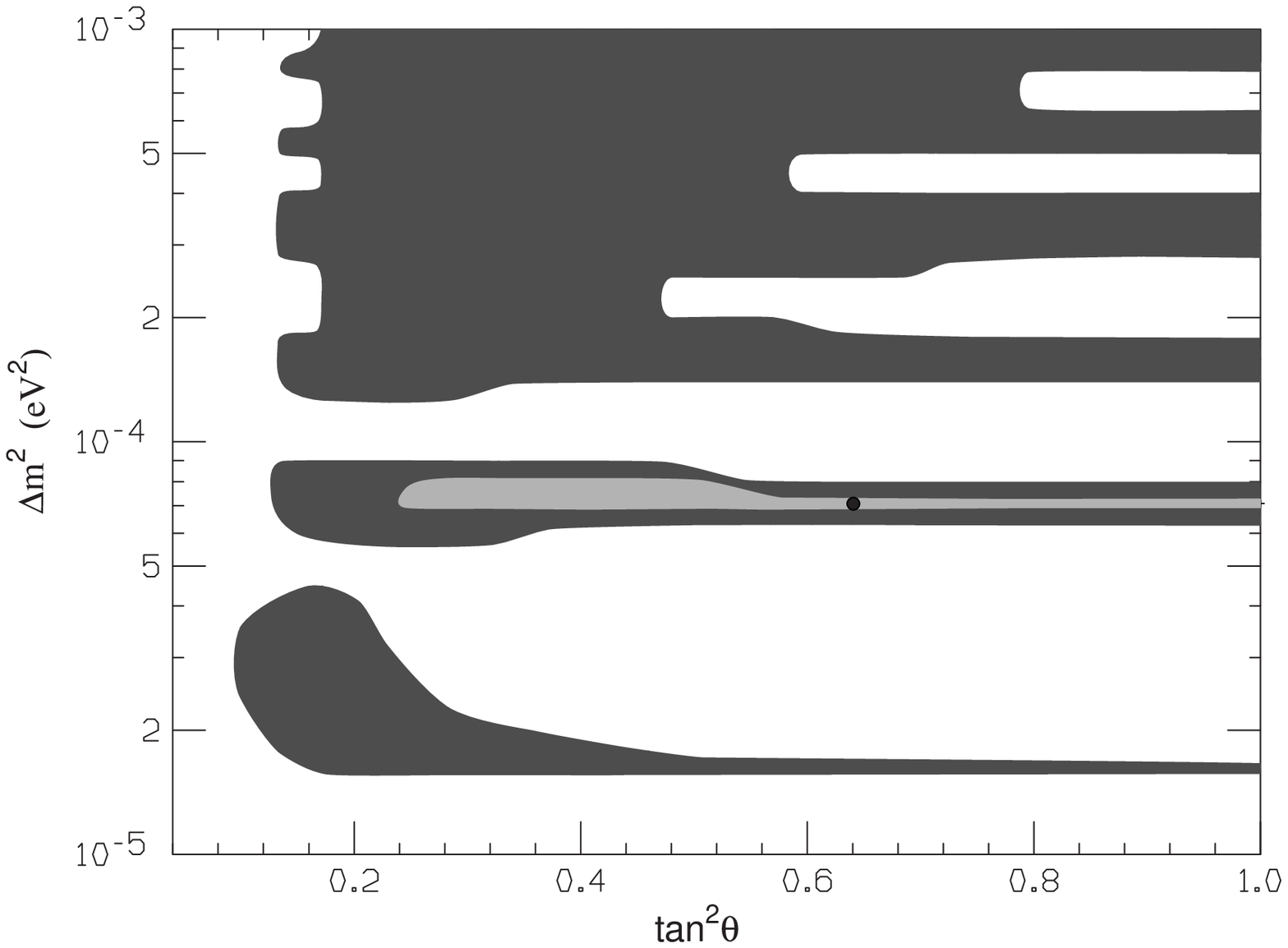,width=14cm,height=14cm}}
\medskip
\caption{The 1$\sigma$ and 2$\sigma$ allowed regions
from a fit to KamLAND data only. The best-fit point is at
$\Delta m^2=7.1\times 10^{-5}$ eV$^2$
and $\tan^2 \theta=0.64$. The figure is symmetric under reflection
about $\tan^2 \theta=1$.}
\label{fig1}
\end{figure}

We first show the results of an analysis of KamLAND data alone to demonstrate
that our assumption that the live times and efficiencies of all the reactors
are the same does not affect the allowed regions. The $1\sigma$ and $2\sigma$
allowed regions are shown.
The similarities between Fig.~\ref{fig1} and Fig.~6 of Ref.~\cite{kamland}
are convincing after accounting for the fact that we have chosen 
$\tan^2 \theta$ as the abscissa. 
The best-fit solution is $\Delta m^2=7.1\times 10^{-5}$ eV$^2$
and $\tan^2 \theta=0.64$ with $\alpha=1.008$ and $\chi^2=5.57$. 
In the LOW region we
find $\chi^2=19.89$ which
is therefore acceptable only at the $3.4\sigma$ C.~L. (KamLAND quotes 
99.95\% C.~L.~\cite{kamland} which is equivalent to about 3.5$\sigma$).
Note that with solar neutrino data alone, the LOW solution is allowed at the 
99\% C.~L. or about 2.6$\sigma$~\cite{Barger:2002iv}. Thus, KamLAND data
 already constrains the LOW solution more than solar data.

\begin{figure}[ht]
\centering\leavevmode
\mbox{\psfig{file=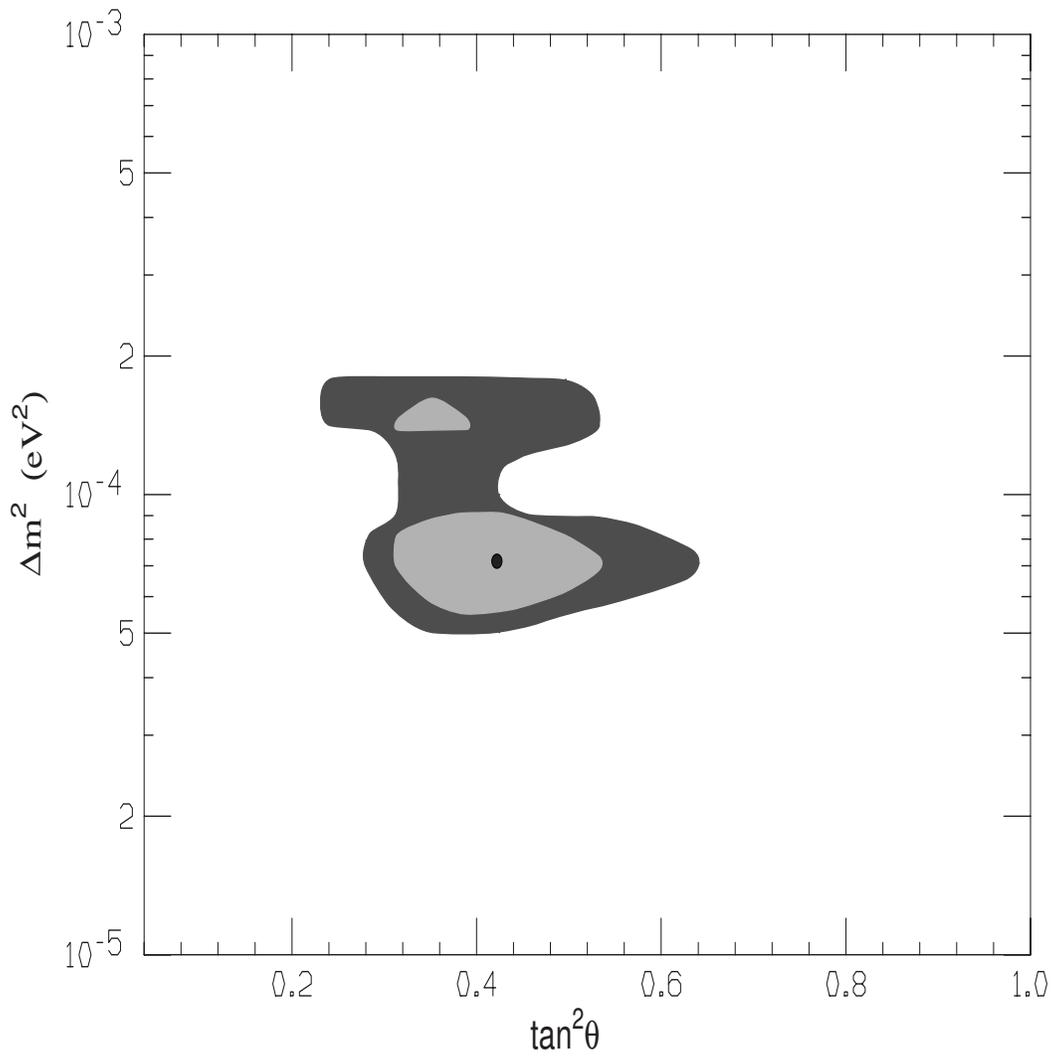,width=14cm,height=14cm}}
\medskip
\caption{The 2$\sigma$ and 3$\sigma$ allowed regions from a combined fit
to KamLAND and solar neutrino data. The best-fit point is at $\Delta
m^2=7.1\times 10^{-5}$ eV$^2$
and $\tan^2 \theta=0.42$.}
\label{fig2}
\end{figure}

In Fig.~\ref{fig2} we show the $2\sigma$ and $3\sigma$ 
allowed regions from a combined analysis of
KamLAND and solar neutrino data. The best-fit solution moves to 
$\Delta m^2=7.1\times 10^{-5}$ eV$^2$
and $\tan^2 \theta=0.42$ with $\alpha=0.994$ and $\chi^2=57.08$. 
The best-fit point
in the LOW region has $\chi^2=79.78$ thereby implying that the LOW solution is
allowed only at $4.4\sigma$.

We conclude that the LMA solution is unique at the $4.4\sigma$ C.~L. A
precise determination of the oscillation parameters is now only a 
matter of time~\cite{barger}.

\vskip 0.4in
\noindent
{\it Acknowledgements.}  This work was supported in part by the
U.S. Department of Energy under grant
Nos.~DE-FG02-91ER40676 and 
~DE-FG02-95ER40896, and in part by the
Wisconsin Alumni Research Foundation.

\clearpage

\end{document}